Seema B. Hegde*, B. Satish Babu and Pallapa Venkataram

# A Cognitive Theory-based Opportunistic Resource-Pooling Scheme for Ad hoc Networks

**Abstract:** Resource pooling in ad hoc networks deals with accumulating computing and network resources to implement network control schemes such as routing, congestion, traffic management, and so on. Pooling of resources can be accomplished using the distributed and dynamic nature of ad hoc networks to achieve collaboration between the devices. Ad hoc networks need a resource-pooling technique that offers quick response, adaptability, and reliability. In this context, we are proposing an opportunistic resource-pooling scheme that uses a cognitive computing model to accumulate the resources with faster resource convergence rate, reliability, and lower latency. The proposed scheme is implemented using the behaviors-observations-beliefs cognitive model, in which the resource-pooling decisions are made based on accumulated knowledge over various behaviors exhibited by nodes in ad hoc networks.

**Keywords:** Mobile ad hoc networks (MANETs), resource pooling, opportunistic computing, cognitive agents, belief.

**2010 Mathematics Subject Classification:** 68.

## 1 Introduction

Amid the extensive use of resource-rich mobile devices with increased abilities to compute and communicate, ad hoc network computing models suffers from ineffective usage of resources. Efficient pooling of available resources can address this issue. Resource pooling deals with the grouping of resources (assets, equipment, personnel, effort, and so on) to maximize the advantages and/or minimize the risk of the users [30]. Resource pooling can create a sense of continuous or infinite resource availability under a volatile network topology. It allows various computing devices and services to share their abilities in order to optimize resource usage among a set of cooperating nodes. Resource pooling is required whenever a node, during communication, experiences scarcity of resources for the successful completion of the task.

In ad hoc networks, mainly mobile ad hoc networks (MANETs), resource pooling will take place among nodes within a radio distance [13]. However, for the collaboration, radio distance may not be the sole criteria. Other criteria [1, 22, 29, 34, 40] can also be considered such as contact frequency, latency, trust, history, node mobility, and node density.

Even then, the efficiency of resource pooling is hindered by the present approaches. These approaches lag in order to cope with the dynamic and adaptive behavior of MANETs due to volatile topology [19, 21] and lack of mobility awareness of the resources. Therefore, a scheme is required to overcome the hurdles faced in

*Corresponding author: Seema B. Hegde,** Siddaganga Institute of Technology, Department of Computer Science and Engineering, Tumkur, India, e-mail: seemab_hegde@yahoo.co.in. http://orcid.org/0000-0002-4495-352X
**B. Satish Babu:** Siddaganga Institute of Technology, Department of Computer Science and Engineering, Tumkur, India
**Pallapa Venkataram:** Indian Institute of Science, Department of Electrical Communication Engineering, Protocol Engineering Technology Unit, Bangalore, India the active collaboration of resources. The design of the scheme should follow the principle of communicating under intermittent seamless connectivity and can be achieved through opportunistic networks [18]. The scheme needs to adopt a context-based approach for collaborating the resource when the nodes opportunistically come in contact in order to cope with the volatility of the environment. Most of the existing schemes are reaction based and lacks the sensitivity toward forecasted decision while gathering the resource availability information due to incomplete knowledge of its open, dynamic environment. The behavior-based decision-making systems are useful in building a robust and reliable system in the dynamic environment [6], as they should be able to initiate resource-pooling process by evaluating its environment. The cognitive behavior can be a design metaphor when developing such a strong opportunistic model, as it should be able to learn and take a decision toward reaching the end goal [32]. Such a system can provide a dynamic and reliable resource-pooling solution. As an opportunistic computing (OC) model, along with cognitive agents (CAs), includes these desirable features, it appears to be a silver lining toward the implementation of such a design to overcome some of the inefficiencies in resource pooling.

## 1.1 Opportunistic Computing

OC is a type of pervasive computing and a nomenclature, and thus, it takes the benefits of randomness and uncertainty in the environment in order to establish communication between pairs of diverse devices and applications opportunistically. It exploits the transient unpredicted contacts to cooperate and share each other's content, resources, and services [14, 28]. The opportunity comes whenever the nodes move in close contact, i.e. within the single hop, direct communication, selected intermediate, source, or destination node. Such nodes are said to be in opportunistic contact if they share a common goal, common resources, and similar context activities such as same background or affinity.

For example, let us consider a social *self-organizing pervasive network* having several nodes with the orchestration of resource such as processing power, memory, software application, sensors, cameras, and so on, as shown in Figure 1. In a service-oriented computing scenario where Alice is a service manager who needs to answer a number of queries from various end users, there is increased waiting time for the queries in the queue. With OC, Alice can get the queue management be taken care by trusted colleagues Bob and James, who came as her opportunistic contact during past intervals. Hence, an OC model will pool up the resources to make them seamlessly available for applications. By implementing the OC model based on cognitive theory, the beliefs [31] on the resource availability and nature of nodes can be obtained. This model improves operational robustness, as it can efficiently provide intelligent pre-failure avoidance and reduce the operational cost. The CAs can model the real world [9]. They can provide information on the availability of various resource with minimum resource management overhead. Thus, this model is proposed to be implemented using CAs.

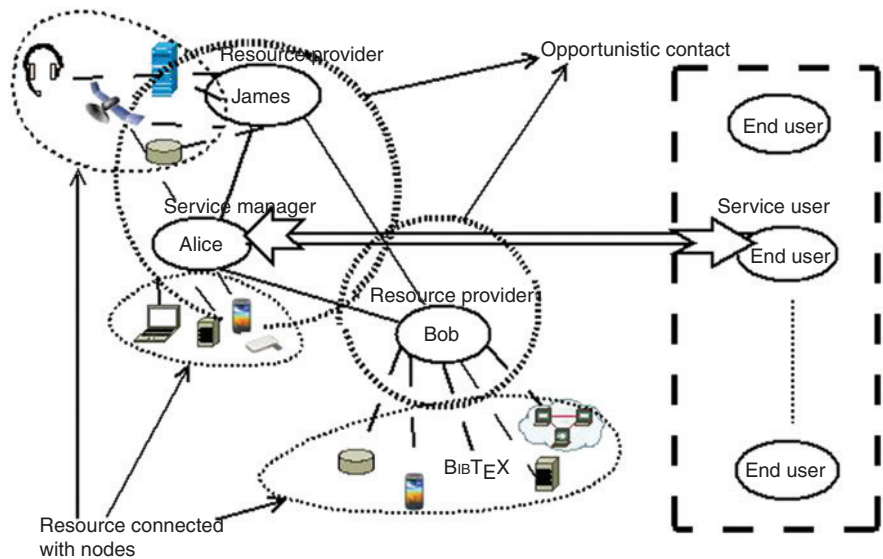

**Figure 1:** Logical architecture of opportunistic service-oriented computing scenario.
The scenario consists of orchestration of resources, where Alice's, the service manager, task of queue management is taken up by trusted colleagues Bob and James on opportunistic contact.

### 1.2 Cognitive Agents

CA [12, 17] is an intelligent software agent. It is an autonomous and responsive entity that reflects the state of information of the environment, which is represented using cognitive terms. Similar to human perceptual experience and understanding of the surrounding environment, CAs can provide an immediate or controlled response to the changes in the environment. Owing to its nature of humanness, CAs can be cloned, dispatched, and activated or deactivated based on changes in the environment. An agent is created, unchanged, and executed on the agent platform. An agent platform comprises an agent server, agents, agent transport mechanism, and an agent execution environment. Usually, inter-agent communication will be done using message passing. The mobile agent code is platform independent, and it can be cloned and executed at any remote node in the heterogeneous network environment. The agent can update its information base while interacting with other agents in the environment. Thus, an agent-embedded model will be more flexible, adaptive, very intuitive, and user-friendly [10, 35]. CAs are mainly characterized by self-learning, self-improving, and self-instructing ability [20, 32], thus reducing network overhead. CAs are suitable for adaptive, dynamic, context-based behaviorism, decision making, human-mobile network interaction system [2, 26, 27], and so on.

### 1.3 Proposed Cognitive Agents-Based Opportunistic Scheme for Resource Pooling

The proposed opportunistic scheme works with CAs. The instance of mobile cognitive agents (MCA) runs on every registered node in ad hoc networks. For a node in need of a resource to execute a task, it should collaborate with other devices that come to its opportunistic contact and that have the required resources for which the MCA generates the belief about the resource availability, based on the observations generated by various observed behaviors on the node. It will communicate with all reachable MCAs through a request (REQ) and reply (REP) model to obtain the beliefs from the neighboring nodes. The belief record stored these beliefs and updated them periodically.

### 1.4 Organization of the Article

The remainder of the article is organized as follows: Section 2 gives related work on essential resource-pooling approaches in MANETs with their working principle and limitations. Section 3 provides definitions of terminology used in the proposed resource-pooling schemes. Section 4 discusses the proposed CA-based belief formulation scheme for resource pooling and its architectural functionality. Section 5 gives the performance analysis of the scheme. Section 6 presents the experimental scenario with the results. Finally, the article concludes with future work.

## 2 Related Works

The resource-pooling architectures are categorized into directory-based, directory-less, and hybrid schemes [15, 24], either with an overlay or without an overlay depending on underlying application scenario. The literature survey of various resource-pooling techniques is summarized below.

The hierarchical radio resource allocation scheme [3] is a directory-based resource-pooling technique with based on the principle of dynamic resource pooling for device collaboration within a radio distance. It uses various radio access technologies with agents.

The model employed for resource pooling includes different stages:

- The node, seeking the resources, sends the *resource request* with all the necessary information through a common channel upstream the MANET until it hits the appropriate agent level. On availability, the resource allocation reply comes through the channel through a resource-allocating agent. Onward, the node can use the allocated radio spectrum.
- The *resource reuse* achieves optimal resource usage with the mapping of the reuse factor through the signal interference ratio based on wireless capacity. All the nodes in the MANET will be monitoring the interference ratio. If the ratio goes below the threshold level $Th_r$ for the time more than $t_r$, then the radio resource will be reinitiated for the new process.
- The agents will trigger *synchronization* among all the nodes for resource accounting purpose and get the child agent resource requirement, thus obtaining reliability in allocation. Then the parent agent sums up each child agent requirement and allocates frequency sizes accordingly.
- With the growth in the use of MANETs in managing the dynamic nature of the network, there is a need for *agent status transfer* to a higher level. The wireless capacity and Euclidean distances from the child nodes are considered in order to determine the suitability of a node for promotion and are given by *agent metric*:

$$\text{Agent metric} = \frac{C}{\sum \sqrt{(X_i - X_{parent})^2 + (Y_i - Y_{parent})^2}}, \quad (1)$$

where $C$ is the wireless capacity of the node under consideration, $(X, Y)$ is the coordinate of the child agent node $i$ that it would serve, and $(X_{parent}, Y_{parent})$ is the coordinate of the prospective node agent to be promoted.

If any node needs a resource, it has to climb through the hierarchical level to reach the required agent that incurs overhead; this is not advisable for MANET-based applications. The dynamic frequency reuse is not optimal, as the relocation is through the master of the cluster instead of the agents.

Reliable server pooling (RSP) [38] is one of the earlier overlay resource-pooling technique that is based on the principle of transparent switching between name domains on failure based on name server technology. The homogeneous servers closer to each other are pooled together and named through entity *name server* (NS) to which *pool element* (resource providers) and *pool users* (pool client) are registered. Switchover mechanism can be further fine-tuned using heartbeat messages [39].

The modeling of server resource pooling was quantified based on the following performance metrics [39]:

- *Session sustainability throughput* (SST): The ratio of the number of sustained sessions to the failed sessions. Sustained sessions are the class of failed session, which will be able to recover from the failure.
- *Switchover efficiency* (SE): The average number of switchover attempts per successful switchover.
- *Session sustainability gain* (SSG): The increase in the system's ability to admit and run a session until completion.

The SSG will quantify the extra overhead incurred by switching to the existing session on the network and thereby indicates the opening of fewer new sessions.

Let $D$ represent the number of session open, $F$ be the failed sessions, and $Y$ be the weighted percentage of those rejected or lost.

Let $p$ and $i$ denote the parameters with and without RSP, respectively.

Let $w_l$ and $w_r$ be the weights the lost and the rejected sessions, respectively. Typically, $w_l > w_r$, then

$$\text{SSG} = \frac{Y^i - Y^p}{y^i}; \quad (2)$$

on optimization,

$$SSG_{max} = \frac{F_d^i}{F_d^i + R_o^i}. \tag{3}$$

To cope with the dynamic topology, the technique performs global synchronization through the peer discovery process and maintains redundant name servers to enhance reliability in pooling. The node has to perform an extensive name hunt to obtain the appropriate name domain that again incurs high overhead and latency. With increasing mobility, the load-balancing performance degrades due to the frequent initiation of peer discovery process.

The flooding technique [8] is an overlay structured resource-pooling technique in high-mobility MANETs based of the principle of optimization to disseminate resource availability information. The technique can dynamically switch among one of the three flooding techniques, i.e. pure flooding, 1 hop, and 2.5 hops, based on the principle of dynamic virtual overlay network [7], to cope with volatility. It comprises an algorithm that forms a set of nodes called broadcast group (BG). One of the nodes in the BG is selected as a leader, called broadcast group leader (BGL), using the parameters $\eta$ and $\mu$. The BGL interact with other group nodes using hit messages to broadcast resource availability information and query message.

The modeling of the algorithm for BGL selection is as follows:

The node $n_x$ will elect the BGL

(i) If $BGL(n_x) = \mu$ (if multiple nodes have it, the node with the smallest address value is elected), where $\mu$ measures the highest link stability value of its neighbors in its neighbor link stability table.
(ii) If no entry exist, then $BGL(n_x) = \eta$ highest of all neighbor nodes, where $\eta$ measures the neighbors link stability and is said to be stable if it is above the set $Stability_{threshold}$.
(iii) $BGL(n_x) = \sum BGL(n_1) + BGL(n_2) + \ldots$.

The dissemination process incurs an overhead because at higher mobility, the look-up efficiency and throughput decrease and the performance degrades.

Efficient resource discoveries (ERD) [37] is a controlled flooding-based resource discovery technique based on the principle of optimizing resource searching and making the most relevant resource available to the nodes within radio distance. It uses a hybrid *push and pull* [36] searching technique with a strategy of spreading high-priority queries obtained based on dynamic ranking.

The algorithm is modeled in various phases, of which during the first phase, *neighbor detection* is performed using a query message $q_i$ consisting of the following parameters:

⟨UID $i$, Source ID, ID of requested resources, Number of hops, Time stamp⟩.

The hello message consisting of following parameters:

⟨Requested node ID, Query message, Time stamp⟩.

An acknowledgment list is represented by $[n_1; q_1; T_1]$, $[n_2; q_2; T_2]$, and $[n_3; q_3; T_3]$, and the acknowledgments in this list are discarded once their lifetime expires.

The second phase performs dynamic ranking of the queries using a Gaussian function [10]:

$$f(t) = \frac{1}{\delta\sqrt{2\pi}} * \exp-\frac{(t-\mu)^2}{2\delta^2}, \tag{4}$$

where $\delta$ and $\mu$ are the constants used to determine the maximum time and threshold time, respectively.

The third phase performs the dynamic ranking of the resources using the previously ranked queries.

The protocol performance declines in a high-mobility and a sparse-node condition like in most MANET applications.

INSIGNIA [11] is an IP-based quality of service (QOS) framework that provides a fast, adaptive service based on a soft-state resource management technique for MANETs. It is modeled architecturally to support fast reservation, responsive restoration, and end-to-end seamless adaptation using the inherent flexibility, robustness, and scalability found in IP networks.

Architecturally, INSIGNIA is designed to have following features:
– It has *adaptive services* that can adapt user sessions to the available level of service without explicit signaling between source–destination pairs with both base QOS and enhanced QOS assurance.
– QOS functionality is decoupled from the routing protocol so that it can be plugging a broad range of routing protocols.
– The framework is based on an *in-band signaling* approach that supports mobility and end-to-end QOS for dynamic environments.
– It maintains the QOS of adaptive flows in mobile ad hoc networks using a *soft-state resource management* approach for the management of reservation in MANETs.
– It adopts multiple restoration techniques to provide reliability.

With the increase in mobility, network dynamics stimulate frequent rerouting, thus increasing traffic overhead.

Using mobile agents for resource discovery [23] is a resource-pooling technique that optimizes resource discovery using an agent-based query processing. Resource discovery is modeled using two types of agents: static and mobile. The static agents will run continuously on each host that will monitor the nodes and generate the resource availability information. The mobile agents will gather this resource information and execute the resource discovery efficiently by reducing the traffic and increasing the scalability.

Resource discovery in peer-to-peer (P-P) networks [16] uses a distributed agent-based approach, i.e. the random walk technique. The indirectly connected graph $G(S, V)$ represents the unstructured P-P network with $S$ nodes, and the $V$ edges are viewed as the ones connecting the peers. Here, to locate the resources, the requesting peer creates the mobile agent that chooses a random walk among the neighbor peers, determines their IP address, and clones itself. The process continues until it finds the appropriate peer or the termination of the random walk. The human immune system form of the self-adaptable regulation model is used to control the number of cloning copies. Along with the difference in the network environment here, the performance of the technique also depends on the form of the random walk employed.

The economy-based opportunistic network model [33] is an auction-based scheduling model for resource utilization. Opportunistic networks possessing heterogeneous devices perform resource discovery by forming the resource grid. Here, the nodes in the network join the resource grid along with the available resource information, creating the grid table. The grid table is maintained and regularly updated, which is used further for scheduling using the second auction-based process. The repeated resource discovery process leading to the reformation of the resource grid becomes expensive with the increasing resources.

Table 1 gives the parameters of our emphasis related to the above-discussed works.

From the review on the above-discussed techniques, we can infer that the dynamic changes in the topology of the MANETs lead to frequent initiation of the resource discovery process. As they incur high overhead, these techniques will fail to provide a dynamic solution for resource pooling. Hence, an approach with a trade-off between a fast response with dynamic pooling and lower overhead is needed.

**Table 1:** Summarization of the resource pooling information of the related works.

| Technique citation | Resource type | Resource information | Dynamic | Reliability | Adaptability | Agent usage |
|---|---|---|---|---|---|---|
| [3] | Radio resource | Allocating agent | Yes | Yes | No | Yes |
| [38] | Servers | Name servers | Yes | Yes | No | No |
| [8] | Heterogeneous class of network resources | BGL | Yes | Yes | No | No |
| [37] | Heterogeneous class of network resources | Gaussian function | No | No | No | No |
| [11] | Heterogeneous class of network resources | Resource management table | Yes | Yes | Yes | No |
| [23] | Heterogeneous class of network resources | Static and mobile agents | Yes | Yes | No | Yes |
| [16] | Heterogeneous class of network resources | Mobile agents | No | Yes | No | Yes |
| [33] | Grid of heterogeneous network resources | Opportunistic networking | Yes | No | No | No |

# 3 Definitions

This section provides the definitions of terminologies used in the article. The proposed scheme generates the belief on resource availability using the behavior–observation–belief cognitive model [5], which uses three cognitive factors, namely behavior, observation, and belief.

## 3.1 Behavior

Behavior quantifies the response to the occurrence of the phenomenon in the network during resource discovery. The proposed scheme will make use of a set of behavior parameters such as velocity of node, frequency of interface, resource mapping history, and so on. The parameters are to be captured within its environment or from the log to produce the behavior. These behavior parameters are quantified as low, moderate, high, sparse, dense, random, or deterministic, and so on. The number of parameters may vary from one behavior to another, and the different combination of the same set of parameters may result in distinct behaviors. For instance, BP = {$BH_1$, $BH_2$, $BH_3$, …, $BH_n$} is the set of behaviors produced during a session based on the set of captured behavior parameters {$bh_1$, $bh_2$, $bh_3$, …, $bh_n$}. The probability of occurrence of an $i^{th}$ behavior $BH_i$ as an outcome of the set of captured behavior parameters {bh} is given by

$$P(BH_i) = \sum_{bh_k \in BH_i} \alpha_{bh_k} \times \text{current value of } (bh_k) / \text{maximum possible value of } (bh_k),$$

where $\alpha_{bh_k}$ represents the weight and is given by

$$\alpha_{bh_k} = \frac{\sum_{k=1}^{n} bh_k}{\sum_{l=1}^{n} \sum_{m=1}^{n} bh_{l,m}},$$

where $bh_k$ is the $k^{th}$ behavior parameter captured and $bh_{l,m}$ is the $l$ number of behavior parameters each captured $m$ number of times.

Some of the behaviors, along with their possible values, are given below:
(i) Behavior velocity (high, medium, or low) is associated with the velocity of the node movement, which represents the mobility of the node.
(ii) Behavior node availability (sparse, moderate, or available) is associated with the contact period of the node, which represents the duration of node availability. Similarly, behavior inter-communication (sparse, moderate, or dense) is associated with the interface period.
(iii) Behavior location tracing (random, semi-deterministic, or deterministic) is associated with the node mobility pattern and represents the accuracy in meeting the node.
(iv) The behavior resource mapping history (poor, moderate, or rich) is associated with the resource availability probability of the node and represents the ability of the node to provide the resources.

## 3.2 Observation

Observation quantifies the act of acquisition and summarization of produced behavior. In the proposed scheme, observations such as the node is static, dynamic, highly dynamic, or high profile (resource-rich), medium profile, or low profile (resource-poor) are confirmed based on the produced behaviors, as shown in Figure 2. For instance, let OB = {$ob_1$, $ob_2$, $ob_3$, …, $ob_n$} be the set of observations obtained from the set of behavior {BP}. The probability of occurrence of an observation $ob_i$ is given by

$$P(ob_i / BH_j) = \frac{\bigcup_j P(BH_j) \times P(BH_j / ob_i)}{\sum_{k=1}^{n} P(BH_j / ob_k) \times P(ob_k)},$$

where $P(BH_j)$ is the probability of occurrence of a behavior $BH_j \in BP$.

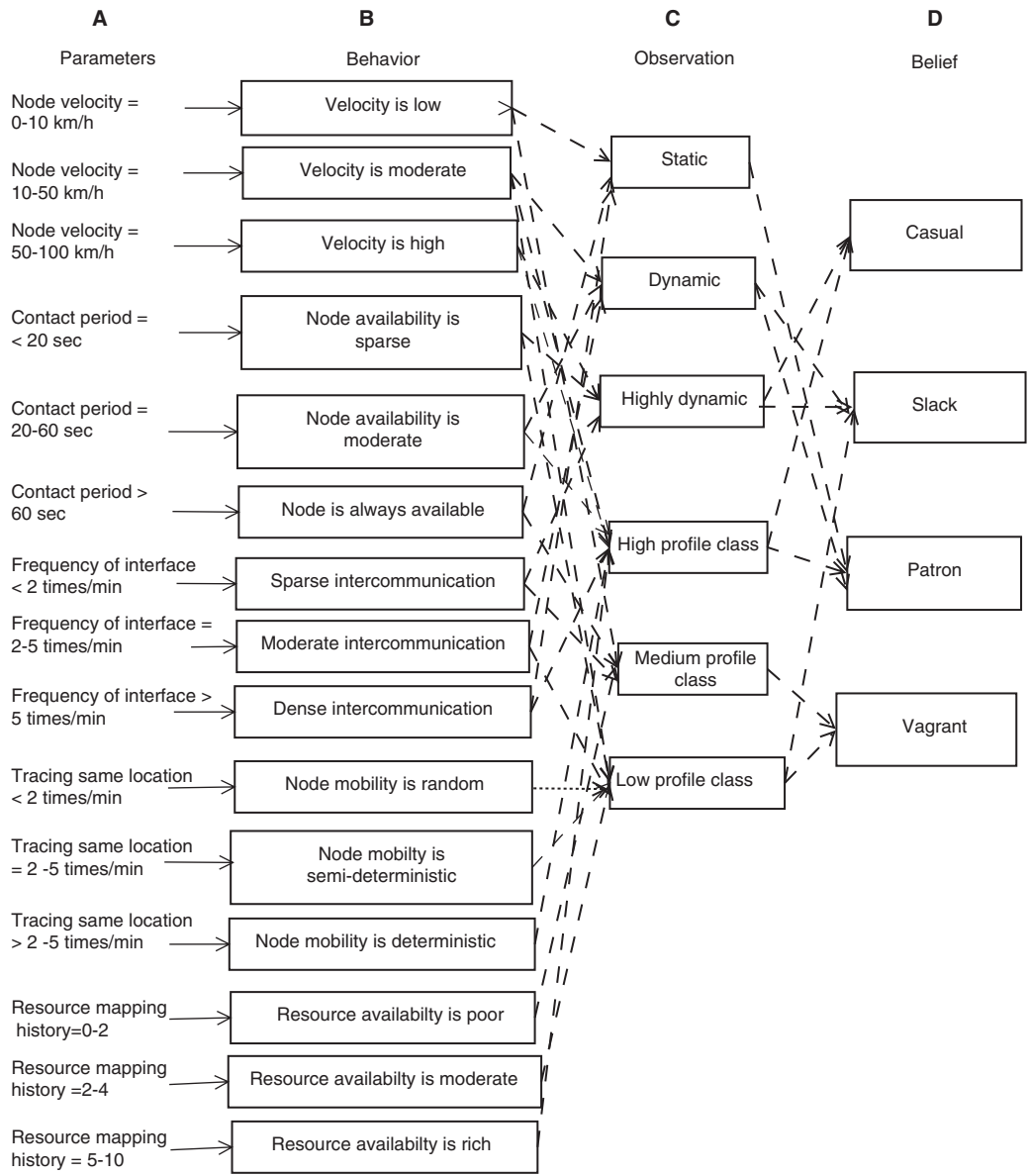

**Figure 2:** An Instance of Belief Generation Over the Resource Availability Using BOB Model.
(A) Behavior parameters captured from external environment or produced by log. (B) Set of behaviors produced based on captured parameters. (C) Set of observation summarized from behaviors. (D) Beliefs generated from the observations.

## 3.3 Belief

Belief quantifies the faith (trust) on a node for resource collaboration based on the interned observations. In the proposed scheme, four types of beliefs namely patron, slack, casual and vagrant are generated, as shown in Figure 2. These beliefs indicate the nature of the node showing resource collaboration is possible or not. For instance, let BL = {$BL_1$, $BL_2$, $BL_3$, …, $BL_n$} be the set of beliefs generated from the set of observations {OB}. The probability of generation of belief $BL_i$ is given by

$$P(\mathrm{BL}_i / \mathrm{OB}_j) = \frac{\bigcup_j P(\mathrm{OB}_j) \times P(\mathrm{OB}_j / \mathrm{BL}_i)}{\sum_{k=1}^{n} P(\mathrm{OB}_j / \mathrm{BL}_k) \times P(\mathrm{BL}_k)},$$

where $P(OB_j)$ is the probability of occurrence of an set of observation $OB_j$. Some of the beliefs and their roles in resource collaboration includes the following: (1) *patron* indicates that the node can actively donate the resources; (2) *casual* indicates that the node is uncertain in providing the resources; (3) *slack* indicates that the node often uses the resources for itself; (4) *vagrant* indicates that the node can not provide any resource support.

## 3.4 Belief Record

Belief record is a tuple ⟨Node ID, Belief generated⟩ that stores the generated beliefs. The belief generated by the MCA running on the host node will be stored as self-belief. Beliefs obtained from all the reachable neighboring nodes through the request and reply model are stored along with their node ID, as shown in Table 2.

# 4 Proposed Cognitive Agents-Based Opportunistic Scheme for Resource Pooling

As MANETs are built under a disconnected environment, to give a continuous sense of resource availability among the nodes, the cognitive agents-based opportunistic scheme for resource pooling (CAOSR) is proposed. The scheme will implement a mechanism in which upon opportunistic contact, a node collaborates with the neighboring nodes for pooling and accumulation of resources using MCA. As in Table 3, the opportunistic contact can be homogeneous, i.e. contact between the same class of nodes, or heterogeneous, i.e. contact between different classes of nodes. We explain how CAOSR works by considering one of the important applications of MANETs, i.e. disaster management.

Figure 3 shows a MANET communication scenario for post-disaster management [4]. Several critical locations around the disaster locations are primarily identified by the disaster management team and starts the rescue and triage operation. The scenario consists of multiple types of nodes such as the node belonging to rescue teams, the node belonging to bystanders, vehicle controllers, navigation controllers, and so on, as defined in Table 4. These nodes are assumed to have diverse levels of capabilities with the resource record (RR) and the MCA installed over them. *Resource record* provides available resource description of a node as ⟨node ID, resource ID, type of resources, …⟩.

The new devices increase in the communication scenario with the addition of bystanders. Within the scenario on detection of the new devices, they will be registered with the navigation controller using a missed call mechanism. Upon dialing the number, the call automatically gets disconnected after two rings. The navigation controller will register the device and send the registration message to the corresponding device on validation, as in Figure 4.

Table 2: Belief Record.

| Node ID | Belief generated |
|---|---|
| Host | Slack |
| ID (neighbor 2) | Casual |
| ID (neighbor 6) | Patron |
| ID (neighbor 3) | Casual |

Table 3: Types of Opportunistic Contact.

| Nodes | Purpose of resource pooling |
|---|---|
| Homogeneous | Share the same class of mapped resource on starvation |
| Heterogeneous | Share the resources that are not available with them on requirement |

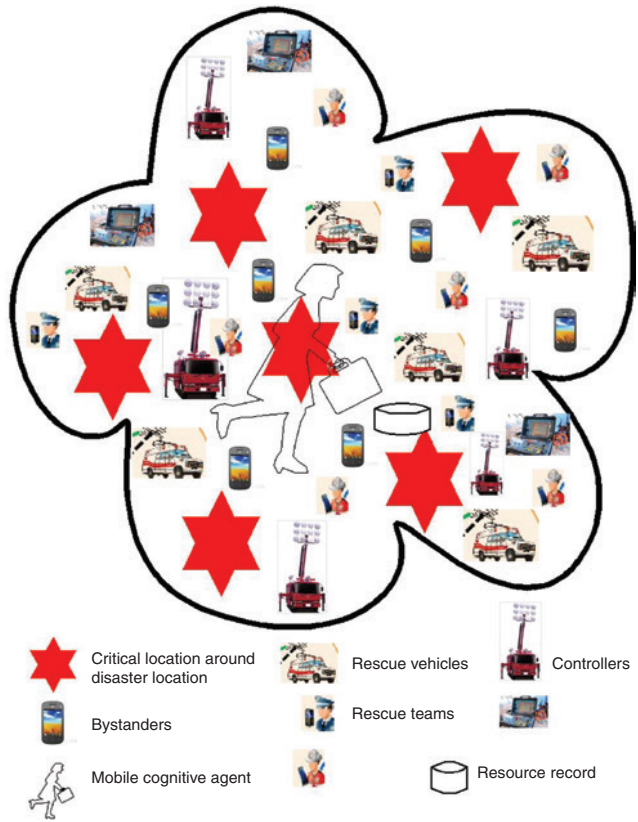

**Figure 3:** MANET Communication Network Scenario for Post-Disaster Management.
The network consist of heterogeneous nodes such as rescue team, bystander, navigation controller, and so on with the RR and MCA installed over them.

**Table 4:** Node Description.

| Node | Description |
| --- | --- |
| Rescue team | A group of trained people allocated with pre-registered devices for faster searching and triage |
| Bystander | The local volunteer individuals registered to assist the rescue team |
| Navigation controller | Registers the new devices and guides the search and triage operation |
| Vehicle controller | Tracks and coordinates the rescue vehicles used for victims triage |

## 4.1 Cognitive Agent Migration

The MCA of the navigation controller, being the registration authority, will clone itself on the newly registered device with the necessary initial handshaking. The copy of the agent from the registration authority will migrate onto the new device, ready to capture the behaviors and formulate the beliefs about resource availability. The MCA on the new device (node) will be available for communication with the MCAs of the other nodes.

The MCA on a node will communicate with reachable MCAs to collaborate with them for pooling of resources. Let us assume some of the nodes mentioned above are mobile (node belonging to rescue team and node of bystanders) and others are static (navigation controller, vehicle controller); some of the nodes are resource rich (nodes belonging to rescue team, navigation controller, and vehicle controller) and others are resource poor (nodes belonging to the bystander). To have a continuous connectivity in such a heterogeneous and volatile condition, the MCAs generate the beliefs and commute them to other MCAs.

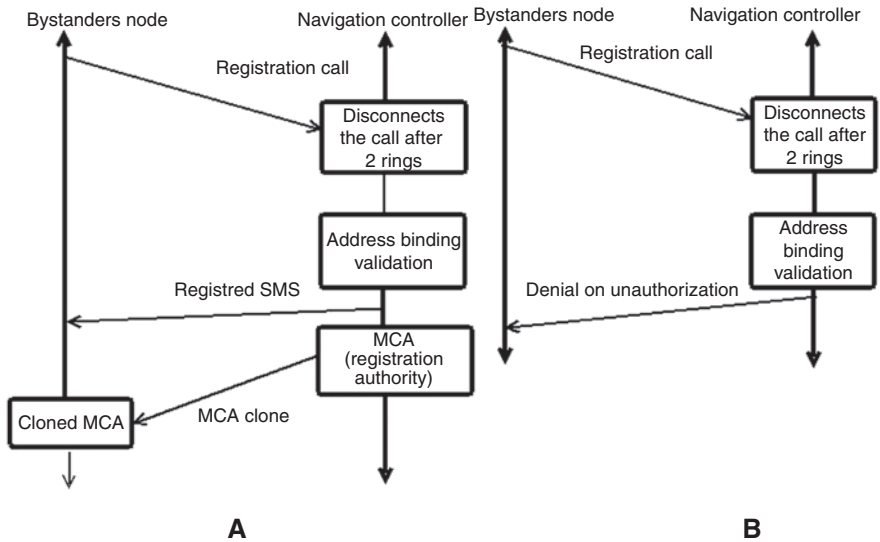

**Figure 4:** Transaction Diagram of Missed Call Mechanism of Registration.
(A) Successful registration. (B) Denial of registration.

## 4.2 The Cognitive Agents-Based Opportunistic Scheme for Resource Pooling

The architecture of the CAOSR, with its functional components, is shown in Figure 5. The MCA runs on the node and communicate with its reachable MCAs through the REQ and REP model. These REQ and REP mainly communicate the beliefs among the nodes. MCA is a footprint mobile agent responsible for formulating the beliefs about the availability of resources on the node. The captured behavior parameters become the input for the MCAs' belief formulator logic. The belief formulator generates the belief about the availability of resources. The MCA stores these beliefs in the belief record along with the beliefs communicated from the neighboring nodes. It also continuously monitors the host nodes RR and proactively finds the demand for resource collaboration (prior to the actual requirement).

## 4.3 Belief formulator

The logic, as shown in Figure 6, is based on the BOB model. The three constructs, namely behaviors identifier, observations generator, and beliefs formulator, are put together to form a belief formulator logic. The belief formulator will formulate the belief over the availability of resources on a node in an ongoing session.

    The behavior identifier gets the captured behavior parameters {bh} as input and using them probabilistically produces the behaviors as shown in Algorithm 1. The probabilistic combination of the various behaviors leads to observations. These summarized observations are stored in the *observation storage* as in Algorithm 2, which are used to generate the beliefs.

    The favorable observations are probabilistically combined to generate any of the four mentioned beliefs. These generated beliefs are stored in the belief record along with the beliefs obtained from the reachable neighboring nodes, which will be updated periodically as in Algorithm 3. The generated beliefs will provide the proactive opportunity for resource collaboration among the nodes, which can be further incorporated as

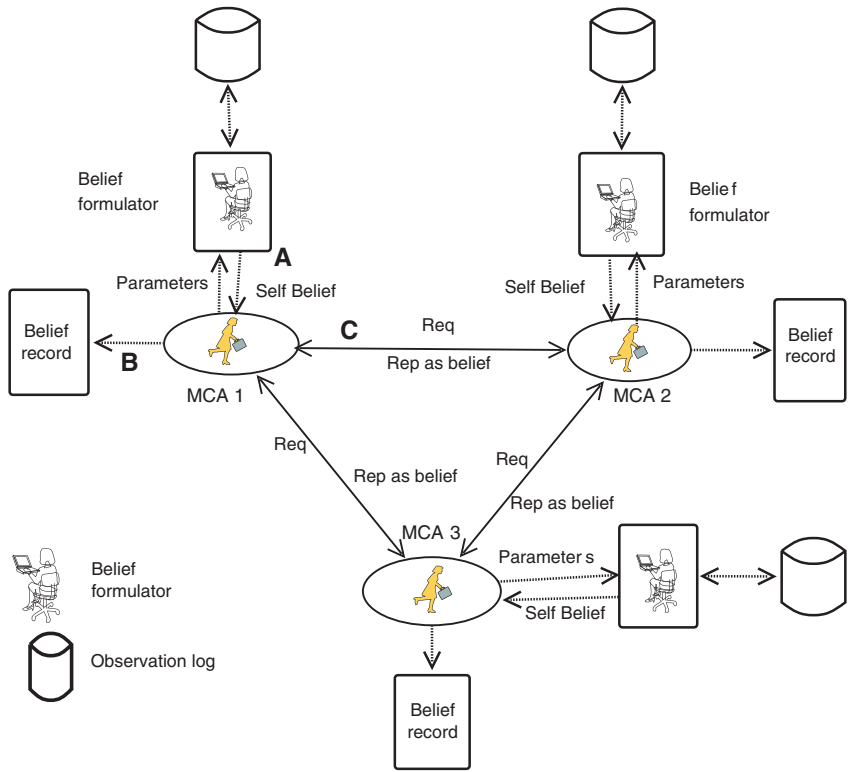

**Figure 5:** The Architecture of CAOSR with MCA as a Main Functional Component.
(A) MCA will pass the captured behavior parameters to the belief formulator and gets back the generated belief called as self-belief. (B) The belief record stores all the collected beliefs. (C) The MCA communicates with all the reachable MCAs through the REQ and REP model.

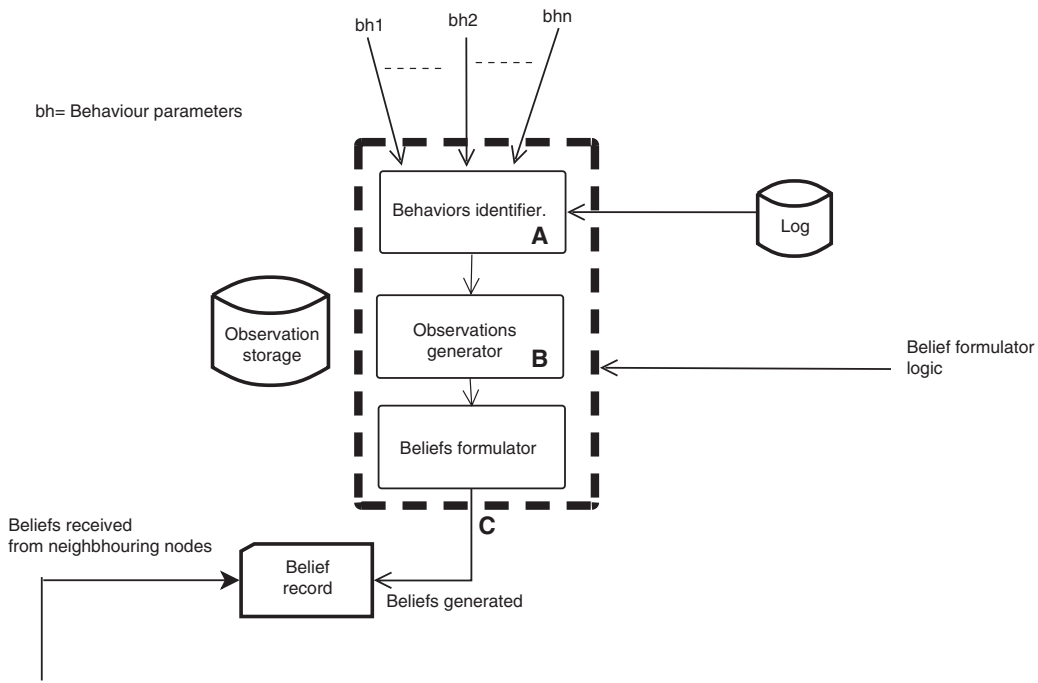

**Figure 6:** The Components of Belief Formulator: Three Constructs.
(A) Behavior identifier: captures the behavior parameters and produce behaviors. (B) Observation generator: summarizes the favorable behaviors into observations. (C) Belief formulator: generates the belief based on observations and stores in belief record.

**Algorithm 1:** Working of Behavior Identifier

1: Begin
2: Initialize the session
3: Initialize the *Log*.
4: **INPUT:** Behavior parameters {$bh_1$, $bh_2$, ..., $bh_n$}.
5: **OUTPUT:** Generated Behavior set BP.
6: Set BP to ø, contains the set of behaviors BH *produced during the session*
7: *while* Not end of session *do*
8:    *Capture the behavior parameters* bh externally and from log.
9:    Pass bh to *Behavior Identifier*.
10:   Compute the weighted $\alpha_{bh_k}$ for $bh_k$
11:   $\alpha_{bh_k} = f(bh_k, bh_{l,m})$
12:   $BH_i^k = f(\alpha_{bh_k}$, current value of ($bh_k$), maximum possible value of ($bh_k$))
13: *end while*
14: *if* BP ≥ 0 *then*
15:   BP = $BH_i \cup$ BP
16: *end if*
17: *Return* BP
18: *Periodically refresh* Log
19: End

**Algorithm 2:** Working of Observation Storage

1: Begin
2: Initialize the *Observation storage*.
3: Obtain behavior BP from *Behavior Identifier*.
4: **INPUT:** Newly generated behavior set BP.
5: **OUTPUT:** Summarized observation set OB.
6: Set OB to ø, contains the set of observations *ob produced from BP*
7: *Pass* BP to observation storage.
8: $\forall ob_i \in OB_i$, *select favorable* $BH_j \in$ BP
9: $ob_i = ob_i \cup BP_j$, $BP_j$ *is the set of favorable behaviors* $BH_j$
10: *while* OB ≥ 0 *do*
11:    OB = $ob_i \in OB_i$ *occurred most number of times during the session*
12: *end while*
13: *Return* OB
14: *Periodically refresh* Observation storage
15: End

a part of routing protocols, where the path establishment/re-establishment phases can be efficiently done with minimum time and link failure.

# 5 Performance Modeling of the Cognitive Agents-Based Opportunistic Scheme for Resource Pooling

The performance of the CAOSR is modeled in terms of reliability in resource availability, rate of convergence, and latency.

## 5.1 Reliability

In CAOSR, reliability ($R$) can be defined as the ability of the MCA on the node to provide the belief on the availability of the resource. The MCA providing the information on the availability of resources depends on

**Algorithm 3:** Working of Belief Generator

---

1: Begin
2: Initialize the *Belief record*.
3: **INPUT:** Newly summarized observation set OB.
4: **OUTPUT:** Generated Belief BL.
5: Set BL to ø, contains the set of beliefs *BL generated from the observations* OB
6: *Summarize the observations* OB *from the observation storage.*
7: *Pass* OB to Belief generator.
8: ∀BL$_i$ ∈ BL, *select favorable* OB$_j$ ∈ OB
9: BL$_i$ = BL$_i$ ∪ OB$_j$
10: *while* BL ≥ 0 *do*
11:   BL = BL ∪ BL$_i$
12: *end while*
13: *Return* BL
14: *if* the REQ & REP model is active, *then*
15: Belief record ⇐ BL$_{Neighbor}$ ⊂ BL
16: *end if*
17: *Periodically refresh* Belief record
18: End

---

the probability of occurrence of resourceful opportunistic contact between the MCAs, $P_{OC}$. The $P_{OC}$ can be obtained by the number of resourceful opportunistic contacts (an opportunistic contact, able to give the resources) between the MCAs, $N_{Roc}$, and the total number of contacts, $N_{tc}$, i.e.

$$P_{OC} = N_{Roc} / N_{tc}, \tag{5}$$

and the total number of contacts is the sum of the number of resource's fewer opportunistic contacts ($N_{OC}$), the number of direct contacts ($N_{DC}$), and the number of resourceful opportunistic contacts ($N_{Roc}$) and is given by

$$N_{tc} = N_{OC} + N_{DC} + N_{Roc}, \tag{6}$$

where $N_{Roc}$ is the sum of the number of homogeneous opportunistic contacts ($N_{hom}$) and heterogeneous opportunistic contacts ($N_{het}$) between the MCAs, i.e.

$$N_{Roc} = N_{hom} + N_{het}. \tag{7}$$

The higher the value of ($P_{OC}$), the more reliable is the scheme, i.e.

$$R \propto P_{OC} \tag{8}$$

## 5.2 Convergence Rate

The convergence rate of the resources ($\xi_{RA}$) is defined as the rate of finding the resource availability and to formulate the beliefs over it. In CAOSR, a high $\xi_{RA}$ is desirable, as the MCA on the node needs to be proactive in finding the resource and formulating the beliefs over it. The $\xi_{RA}$ depends on the mobility of the node ($N_m$), which can be determined by the pause time (the node stops for a duration) ($t_p$), i.e.

$$N_m \propto \frac{1}{t_p}, \tag{9}$$

and the node density ($N_d$), given by

$$\xi_{RA} = \frac{N_d}{t_p} \equiv (N_d \times N_m). \tag{10}$$

The rate of resource convergence increases with shorter pause time (higher node mobility, i.e. nodes become dynamic) and higher node density, with which more MCAs will come into opportunistic contact.

## 5.3 Latency

The latency of the scheme is a function of the overall time, including the convergence time in the formulation of belief, using CAs based on BOB model for resource pooling.

## 5.4 Belief Formulation Time

The belief formulation time is the total time taken in the formulation of a belief, i.e. $T_{tot}$. It includes the time taken by the MCA in formulating the belief $T_{bf}$ and time taken for belief exchange with the neighboring node $T_{bx}$, as shown in Figure 7:

$$T_{tot} = T_{bf} + T_{bx}. \tag{11}$$

## 5.5 Self-Belief Formulation Time

The self-belief formulation time $T_{bf}$ is the sum of the time taken for the accumulation of the behavior parameters to produce the behavior ($T_{bh}$), the observation formulation time ($T_{ob}$), and the belief generation time ($T_{BL}$):

$$T_{bf} = T_{bh} + T_{ob} + T_{BL}. \tag{12}$$

## 5.6 Belief Exchange Time

The belief exchange time $T_{bx}$ is the sum of the time taken by the REQ and REP model to request for the belief and response $T_{req}$ and $T_{rep}$, respectively:

$$T_{bx} = T_{req} + T_{rep}. \tag{13}$$

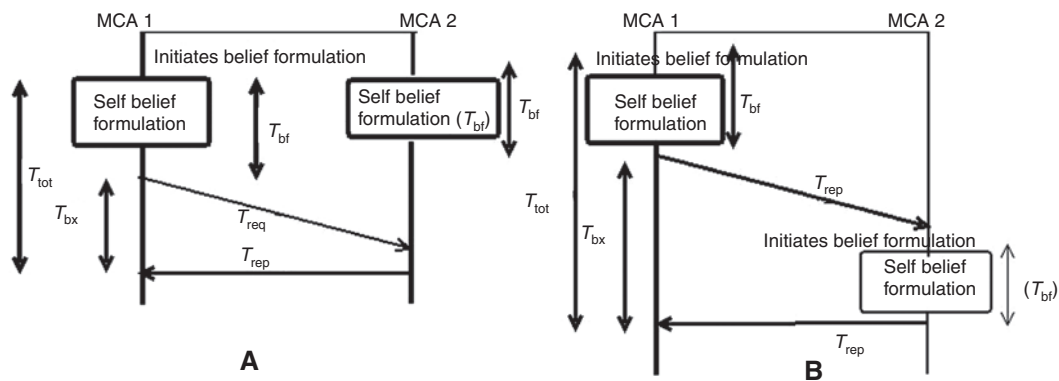

**Figure 7:** Belief Formulation Timing Diagram of MCA1 that Exchanges the Belief with MCA2.
$T_{tot}$ is the total time taken in formulation of belief and is the summation of self-belief formulation time $T_{bf}$ and belief exchange time $T_{bx}$. (A) $T_{tot}$ is small when MCA2 formulates the self-belief before the request from MCA1. (B) $T_{tot}$ is large when MCA2 formulates the self-belief after the request from MCA.

## 5.7 Average Belief Formulation Time

The average belief formulation time $T_{avg}$ gives the average time taken for belief formulation over a node in unit time:

$$T_{avg} = \sum_{i=0}^{n} (\lambda_{bh_i} \times T_{tot}),$$

where $\lambda_{bh_i}$ represents the capturing rate of the behavior parameters $bh_i$ in a session for all values of $i$.

## 5.8 Behavior Capturing Time

The behavior capturing time $T_{bc}$ is the sum of the time involved in capturing the behavior parameters from the external world and from session history:

$$T_{bc} = \sum_{i=1}^{n} T_{bh_i}, \tag{15}$$

where $bh_i$ is the captured behavior parameter. Thus, the behavior generation time $T_{bh}$ is given by

$$T_{bh} = \sum_{i=1}^{n} \sum_{j=1}^{n} m_j \times T_{bc_i}, \tag{16}$$

where $m_j$ is the number of times a behavior parameter $bh_i$ is captured.

The observation generation time can be represented as

$$T_{ob} = \sum_{i=1}^{n} \sum_{j=1}^{n} m_j \times T_{bh_i}. \tag{17}$$

Finally, the belief generation time can be represented as

$$T_{BL} = \sum_{i=1}^{n} \sum_{j=1}^{n} m_j \times T_{ob_i}. \tag{18}$$

## 5.9 Convergence Time

Convergence time $T_{con}$ is the time interval between the request and the location of availability of required resources and is given by

$$T_{con} = T_{rreq} + T_{tra} + T_{tot}, \tag{19}$$

where $T_{rreq}$ is the resource availability request propagation time, $T_{tra}$ is the time taken in tracing the resources through the nodes record, and $T_{tot}$ is the time taken in belief formulation on resource availability.

According to Equation (12), MCA takes $O(n^2)$ time for self-belief formulation as Equations (16)–(18) takes $O(n^2)$ time. The total belief formulation can be done with $O(n^2)$ complexity according to Equation (11). The average belief formulation time will have $O(n)$ complexity.

# 6 Experimental Scenario

An experimental scenario is considered, as shown in Figure 8, to illustrate the working of the CAOSR during disaster management. In an ideal situation, around the location of disaster, despite collapsed networking, intermittent groups of subnets such as Internet links, mobile towers, or communication infrastructure within a building are available. They possess various levels of resource capabilities and device characteristics. These networks provide computational resources such as memory, processor, a communication channel, and software services to offer resource usage, along with the wireless communication devices held by rescue teams.

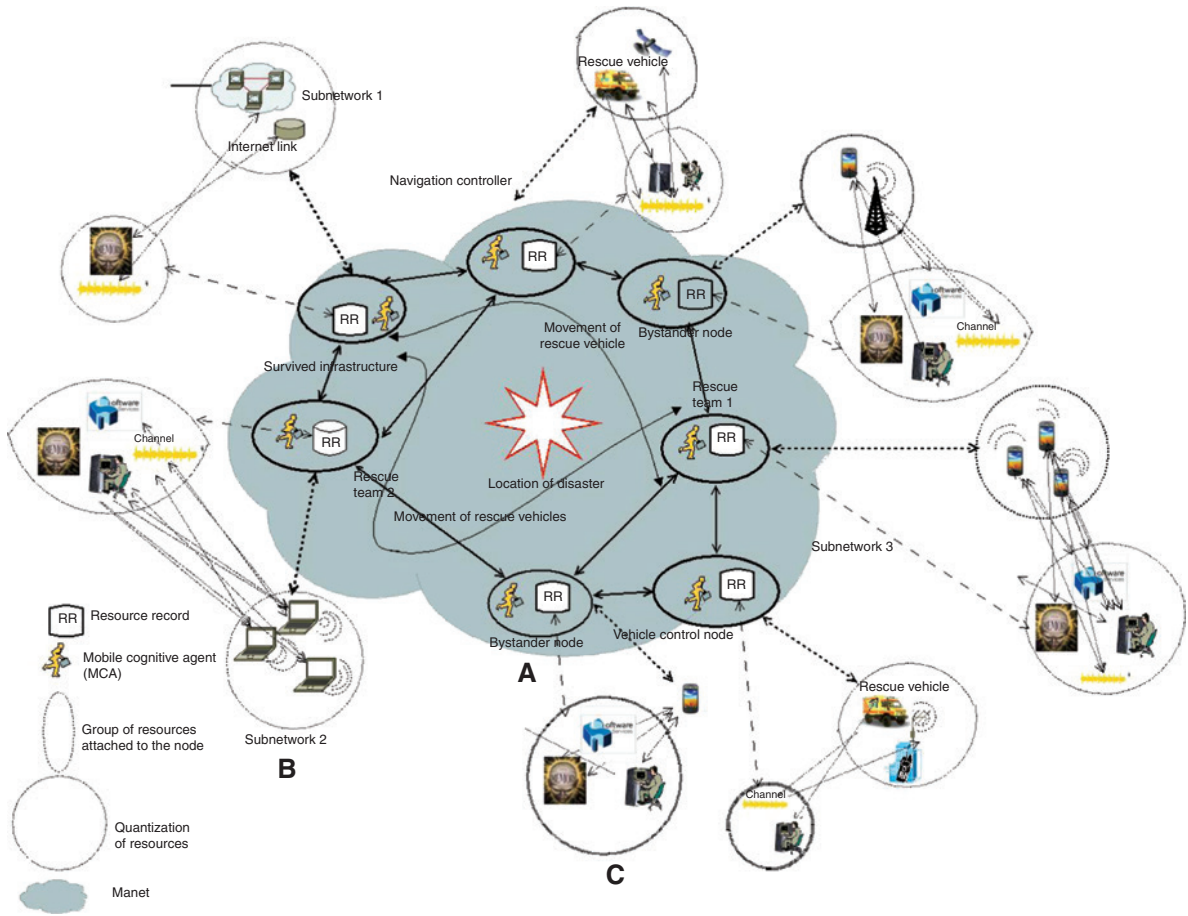

**Figure 8:** An Experimental Scenario of Resource Mapping for Disaster Management.
(A) A MANET of heterogeneous nodes with a diverse level of resource capabilities and device characteristics. (B) The group of devices with resource attached to the node through registration. (C) Quantization of the resources associated with the node stored in the RR.

This model implemented with the CAOSR has events such as *registration*, *quantification*, *mapping*, and *belief generation*. The belief database is set up for more than 200 mobile resources, which involve a mix of mobile devices with random mobility patterns and sporadic data transmission rates. For example, the team of rescue members (nodes) holding the wireless communication devices, such as smartphones, laptops, and so on, will move with a group and have slower unpredictable movement. Thus, there will be much delay in message passing. The rescue vehicles(nodes) will have faster and slightly predictable path, leading to faster message transmissions and so on.

## 6.1 Test Bed Description

The evaluation of CAOSR is done on a test bed of a post-disaster scenario. The pooling of resources is done in a completely or partially collapsed communication infrastructure formed amid the affected area, as shown in Figure 9. The environment comprises 200 wireless mobile nodes embedded with the MCAs designed using agent factory, based on 802.11b configured in ad hoc mode, each with a data rate of 2 Mbps and random node degree.

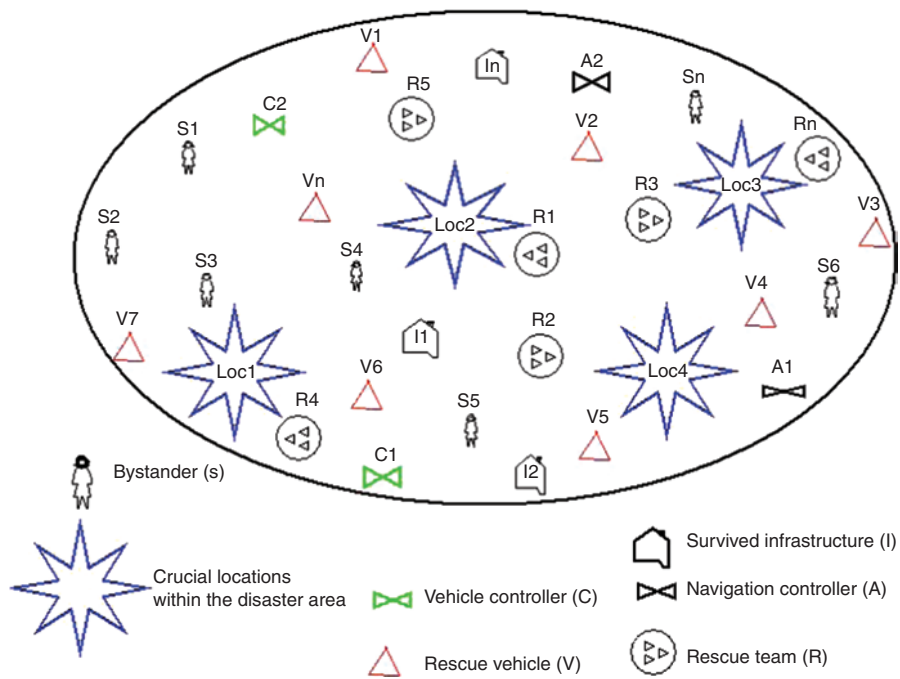

**Figure 9:** Test Bed Setup for Evaluation of Post-Disaster Scenario.

The nodes' mobility is configured to a random waypoint with a communication range of 60 m and pause time ranging from 0 to 150 ms. The nodes are running with the Linux kernel on the IP networking protocol stack that are essential in studying the scenario, of which about two-thirds are mobile nodes such as a rescue vehicle ($V_1$, $V_2$, …, $V_n$) equipped with GPS-enabled IP-based netbook machines with IEEE 1609 connection and with CBR traffic. The nodes belonging to the rescue team ($R_1, R_2, …, R_n$) and bystanders ($S_1, S_2, …, S_n$) have a reduced communication range of 10–30 m and are equipped with notepads, smartphones, and tablets. The remaining nodes are static for all transmission range such as survived infrastructure ($I_1, I_2, …, I_n$), vehicle controllers ($C_1, C_2, …, C_n$), and navigation controllers ($A_1, A_2, …, A_n$) emulated through Mac laptops with wireless cards. The registered devices will attach to the node's resource pool and be mapped with its class of resource and will then update the RR. The test bed gives various possibilities of a node coming in opportunistic contact (nodes in the same direction, toward same locations, executing the same task) with other registered nodes for resource pooling. It uses the RFC 2501 [25] mobile IP networking wireless protocol for node interaction. For instance, $V_5$ coming in opportunistic contact with $R_2$ can share the communication channel required for belief exchange.

Initially, only the surviving infrastructure (static nodes) are positioned around the affected location. With time, the number of nodes increases due to the availability of the disaster management teams and bystanders helping the triage. The plot given in Figure 10 shows the increase in the number of nodes with time. With the increase in node density around the affected location, a resourceful contact opportunity increases.

The plot in Figure 11 shows both homogeneous and heterogeneous opportunistic contacts. It can be observed that the number of heterogeneous opportunistic contacts are more than the number of homogeneous contacts. With the probability of having more heterogeneous opportunistic contacts than homogeneous, a node gains the benefit of using more resources than it has. Thus, a better reliability can be seen in the resource availability. The plot in Figure 12 shows the dependency of reliability over opportunistic contacts. The increase in the percentage of resourceful opportunistic contact will improve the reliability of the scheme.

The plot given in Figure 13 shows the influence of opportunistic contacts on resource availability. The nodes on opportunistic contact can coordinately determine the availability of resources, which forms a medium for dispersion of information throughout the affected area. An agent program running on an MCA captures the required variables to determine the resource availability. Therefore, with an increase in opportunistic contact, the percentage of resource availability also increases.

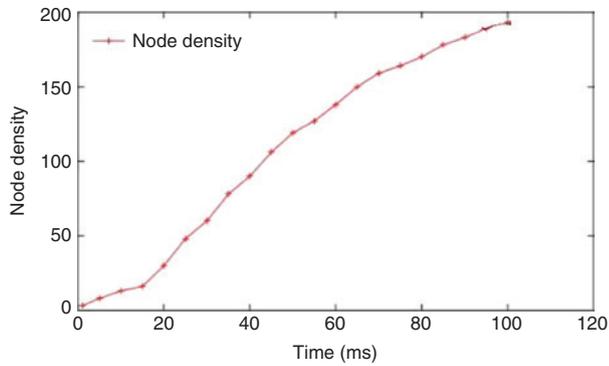

**Figure 10:** Number of Nodes vs. Time (in ms).

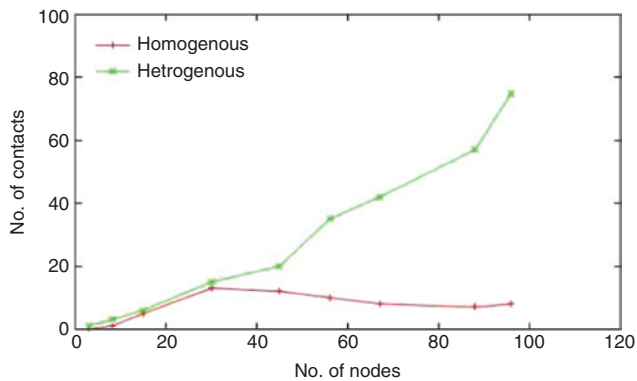

**Figure 11:** Number of Contacts vs. Number of Nodes.

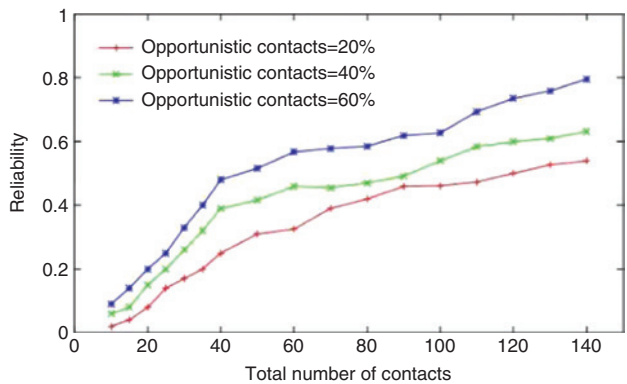

**Figure 12:** Reliability vs. Total Number of Node Contacts.

The plot in Figure 14 describes the effect of node density on the generation of four classes of beliefs. With an increase in node density, resource availability increases and the agent will be able to generate beliefs with reduced belief formulation time. The patron class of beliefs dominates, and the slack class of beliefs degrades with time. This result reflects that more resource can be pooled up with opportunistic contact to manage the critical scenarios dynamically and with reliability.

The plot given in Figure 15 shows the effect of pause time on the resource availability convergence rate. It can be observed that a shorter pause time of the scenario, i.e. higher node mobility, leads to increased resource availability convergence rate. The higher mobility rate of the nodes provides more opportunistic contacts, which helps in faster resource location and hence higher convergence rate. Thus, at higher mobility, the network becomes more dynamic.

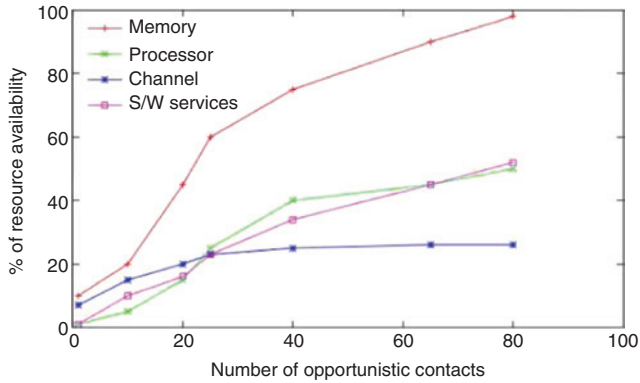

**Figure 13:** Resource Availability vs. Opportunistic Contact.

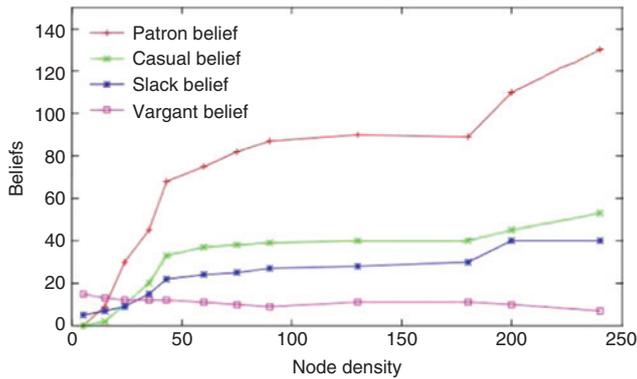

**Figure 14:** Belief vs. Node Density.

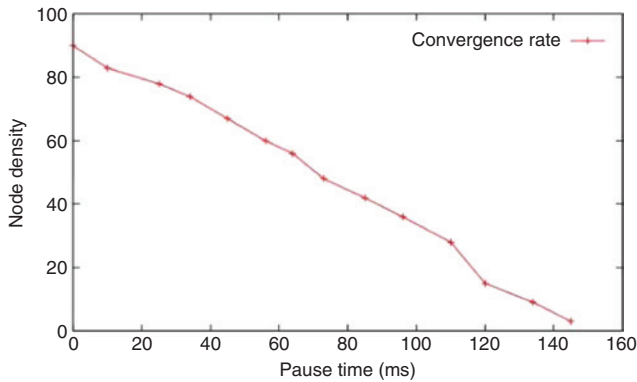

**Figure 15:** Node Density vs. Pause Time.

Figure 16 shows the plot of the percentage of belief formulation failure rate with time. As the resource availability increases, the process of belief formulation is to be increased at the same rate. However, the resource availability convergence rate is also dependent on the mobility of the nodes within the application scenario. Initially, the mobility will be less due to the collapsed pre-existing infrastructure. Therefore, the number of opportunistic contacts will be fewer. Thus, the failure rate is high at the beginning, and with time, it will decrease due to the increase in mobility.

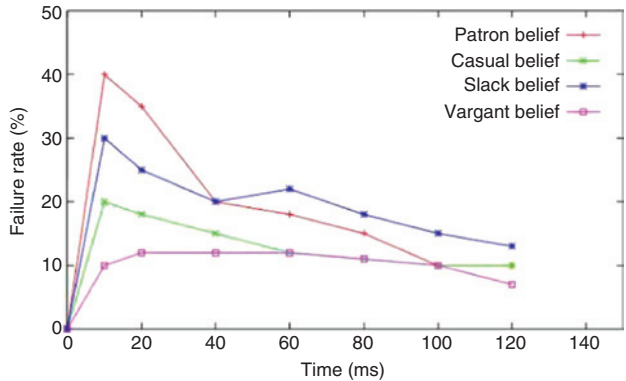

**Figure 16:** Failure Rate (in %) vs. Time Interval.

# 7 Conclusion

This article presents a novel scheme of resource pooling for ad hoc networks to cope with its volatile topology. The scheme uses CA-based OC principles to pool the resources with faster convergence rate and reliability. It is implemented based on the BOB model for a disaster management scenario. The scheme used works close to human cognitive behavior in determining the availability of the resources and is therefore an efficient approach for resources pooling. The scheme can be further extended to provide a dynamic and adaptive routing for resource accumulation in a disconnected environment. The mobility aspect of the MCA should be explored further in future work.

# Bibliography

[1] A. Aggarwal, K. Gopal and A. Kumar, Lightweight trust aggregation through lightweight vibrations for trust, in: *IEEE Conference on Contemporary Computing*, pp. 421–426, 2013.
[2] N. G. Aghaee and T. I. Oren, Effects of cognitive complexity in agent simulation: basics, in: *Proceedings of the Summer Computer. Simulation Conference*, 2004.
[3] A. H. Aghvami, O. Holland and Q. Fan, A dynamic hierarchal radio resource allocation scheme for mobile adhoc networks, *IEEE Trans. Commun.* (2004), 1005–1010.
[4] K. Ahmed and S. Majid, Post-disaster communications: a cognitive agent approach, in: *7th International Conference on Networking*, 2008.
[5] B. S. Babu and P. Venkataram, Cognitive agents based authentication & privacy scheme for mobile transactions (CABAPS), *Comput. Commun.* **31** (2008), 4060–4071.
[6] R. D. Beer, Framing the debate between computational and dynamical approaches to cognitive science, *J. Behav. Brain Sci.* **21** (1998), 630.
[7] L. Bernardo and R. Oliveira, Technical Report, 2005.
[8] L. Bernardo, P. Pinto and R. Oliveira, flooding techniques for resource discovery on high mobility MANETs, in: *Int. Workshop on Wireless Ad-hoc Networks*, London, UK, 2005.
[9] C. Bradley, Cognitive agents interacting in real and virtual world. http://act-r.psy.cmu.edu/wordpress/wp-content/uploads/2012/12/622SDOC4694.pdf.
[10] J. Bradshaw, *Software agents*, AAAI Press, California, 2000.
[11] A. T. Campbell, G.-S. Ahn, S.-B. Lee and X. Zhang, INSIGNIA: an IP-based quality of service framework for mobile ad hoc networks, *J. Parallel Distrib. Comput.* **60** (2000), 374–406.
[12] C. Castelfranchi, Guaranties for autonomy in cognitive agent architecture, intelligent agents: theories, architectures and languages, *LNAI* **890** (2005), 56–70.
[13] C. M. Chin, M. L. Sim and S. Olafsson, Mobile adhoc networks, in: *Encyclopedia of Mobile Computing and Commerce*, chapter 70, 2656, China, pp. 424–428, 2007.
[14] M. Conti, Opportunities in opportunistic computing, *IEEE Commun. Mag.* **43** (2010), 42–50.
[15] D1.2: services models, protocols and security specification, collaborative project with thematic priority: future Internet experimental facility and experimentally-driven research, 2012.

[16] J. Gaber and M. Bakhouya, Mobile agent-based approach for resource discovery in peer-to-peer networks, *LNAI* **4461** (2008), 63–73.
[17] M. N. Huhns and M. P. Singh, Cognitive agents, *IEEE Internet Comput.* **2** (1998), 87–89.
[18] C.-M. Huang, C.-Z. Tsai and K.-c. Lan, A survey of opportunistic networks, *IEEE Dig. Explor.* (2008), DOI 10.1109/WAINA.2008.292.
[19] B. Karaoglu, *Efficient use of resources in mobile ad hoc networks*, Ph.D. thesis University of Rochester, New York, 2013.
[20] D. B. Lange and M. Oshima, *Programming and deploying Java mobile agents with Aglets*, Addison-Wesley, USA, 1998.
[21] Y. Lim, The limit of adhoc networks to commercial success. http://www.cs.rutgers.edu/~rmartin/teaching/fall04/cs552/papers/017.pdf. Accessed 24 September, 2004.
[22] B. L. Mark, M. Hejmo, R. K. Thomas and C. Zouridaki, Robust cooperative trust establishment for MANETs, in: *Proceedings of the 4th ACM Workshop on Security of Ad Hoc and Sensor Networks, Alexandria, VA*, pp. 23–34, 2006.
[23] K. A. Mcartney, N. Migas, W. J. Buchnan, Mobile agents for routing, topology discovery, and automatic network reconfiguration in ad-hoc networks, in: *10th IEEE Int. Conference and Workshop on the Engineering of Computer-Based Systems*, 2003.
[24] A. N. Mian, R. Baldoni and R. Beraldi, A survey of service discovery protocols in multihop mobile ad hoc networks, term paper, *IEEE CS, Pervasive Comput.* (2009), Department of Computer and Information Science and Engineering, University of Rome La Sapienza.
[25] Mobile ad hoc networking (MANET): routing protocol performance issues and evaluation considerations. https://tools.ietf.org/html/rfc2501. Accessed January, 1999.
[26] C. A. Ntuen, An agent-based model simulation of multiple collaborating mobile ad hoc networks (MANET), in: *16th Collective International Command, Control Research and Technology Symposium*, 2009.
[27] J. R. Olson and M. Olson, The growth of cognitive modeling in human computer interface, *J. Human Comput. Interact.* **5** (2005), 221–265.
[28] A. Passarella, F. Delmastro and M. Conti, Mobile service platforms based on opportunistic computing: the SCAMPI Project, *ERCIM News* **93** (2013). http://ercim-news.ercim.eu/en93/special/mobile-service-platforms-based-on-opportunistic-computing-the-scampi-project. Accessed April, 2013.
[29] K. Ponmozhi and R. S. Rajesh, Adhoc resource binding in MANETs, *Proc. Int. J. Comput. Appl.* **37** (2011).
[30] Pooling (resource management). http://en.wikipedia.org/wiki/Pooling_(resource_management). Accessed May, 2009.
[31] A. Postlewaite and O. Compte, Belief formation, *PIER Working Paper* (2012), DOI: 10.2139.
[32] C. Preist, M. Luck and P. Mcuberney, *A manifesto for agent technology: towards next generation computing*, Acution Agent Multi-AG, 203–252, Kluwer Academic, USA, 2004.
[33] B. Saranya and V. Vetriselvi, Utilization of resources effectively by using an economic based model in opportunistic networks, in: *Int. Conference on Computer Communication and Informatics*, 2014.
[34] H. Shen, H. Zhang and K. Chen, Leveraging social networks for P2P content-based file sharing in disconnected MANETs, *IEEE Trans. Mobile Comput.* **13** (2014), 235–249.
[35] M. Wooldridge, Agent-based software engineering, *Proc. IEEE Software Eng.* **144** (1997), 26–37.
[36] J. Wu and Q. Yuan, A dynamic Voronoi regions-based publish/subscribe protocol in mobile networks, in: *Proceedings of IEEE Infocom*, 2008.
[37] J. Wu, A. Srinivasan and R. Thanawala, Efficient resource discovery in mobile ad hoc networks, communications, in: *IEEE Int. Conference*, 2009.
[38] J. Zheng, M. U. Uyar, M. A. Fecko and S. Samtani, Performance study of reliable server pooling, in: *IEEE Int. Symp. Network Comput. Appl.*, Cambridge, MA, 2003.
[39] J. Zheng, M. U. Uyar, M. A. Fecko and S. Samtani, Reliable server pooling in highly mobile wireless networks, in: *Proc. IEEE Int. Symp. Comput. Commun*, Turkey, 2003, DOI: 10.1109/ISCC.2003.1214188.
[40] B. Zhou, G. Yang, J. Chen, L.-J. Chen, M. Gerla, S. Das and Y.-Z. Lee, Dealing with node mobility in ad hoc wireless network, in: *Formal Methods for Mobile Computing* Vol. 3465, pp. 69–106, Computer Science, USA, 2005.